\documentclass[journal,draftclsnofoot,onecolumn,12pt]{IEEEtran}

\usepackage{graphicx}
\usepackage{subfigure}
\usepackage{multirow}
\usepackage{array}
\newcolumntype{P}[1]{>{\centering\arraybackslash}p{#1}}
\newcolumntype{M}[1]{>{\centering\arraybackslash}m{#1}}

\usepackage{amsthm,amssymb,amsmath,graphicx,multirow,color,amsfonts}%
\usepackage[update,prepend]{epstopdf}
\usepackage{multirow}
\usepackage[latin1]{inputenc}
\usepackage{tikz}
\usepackage{bbm} 
\usepackage{pdfpages}
\usepackage{multirow}
\usepackage{subfig}
\usepackage{comment}
 \usepackage{graphicx}
\usepackage{multicol}
\usepackage{cite}

\usepackage[justification=centering]{caption}
\usepackage{textcomp}
\usepackage{psfrag}
\usepackage{arydshln}
\usepackage{url}
\usepackage{soul}
\usepackage{graphicx,color}
\usepackage[nolist]{acronym}
\usepackage{algorithm,algorithmic} 

\usepackage{mathtools,lipsum}
\usepackage{cuted}
\usepackage{amsmath}
\newtheorem{proposition}{Proposition} 
\usepackage{mathrsfs}

\usepackage[capitalise]{cleveref}
\Crefname{equation}{Eq.\!}{Eqs.\!}
\Crefname{figure}{Fig.\!}{Figs.\!}
\Crefname{tabular}{Tab.\!}{Tabs.\!}
\Crefname{section}{Section\!}{Sections.\!}

\def\nb0{{\mathbf{0}}}
\def\nb1{{\mathbf{1}}}

\newtheorem{corollary}{Corollary}

\newenvironment{sequation}{
\begin{equation}\small}{\end{equation}
}

\begin{document}
\graphicspath{{./Figures/}}
	\begin{acronym}

\acro{5G-NR}{5G New Radio}
\acro{3GPP}{3rd Generation Partnership Project}
\acro{ABS}{aerial base station}
\acro{AC}{address coding}
\acro{ACF}{autocorrelation function}
\acro{ACR}{autocorrelation receiver}
\acro{ADC}{analog-to-digital converter}
\acrodef{aic}[AIC]{Analog-to-Information Converter}     
\acro{AIC}[AIC]{Akaike information criterion}
\acro{aric}[ARIC]{asymmetric restricted isometry constant}
\acro{arip}[ARIP]{asymmetric restricted isometry property}

\acro{ARQ}{Automatic Repeat Request}
\acro{AUB}{asymptotic union bound}
\acrodef{awgn}[AWGN]{Additive White Gaussian Noise}     
\acro{AWGN}{additive white Gaussian noise}

\acro{APSK}[PSK]{asymmetric PSK} 

\acro{waric}[AWRICs]{asymmetric weak restricted isometry constants}
\acro{warip}[AWRIP]{asymmetric weak restricted isometry property}
\acro{BCH}{Bose, Chaudhuri, and Hocquenghem}        
\acro{BCHC}[BCHSC]{BCH based source coding}
\acro{BEP}{bit error probability}
\acro{BFC}{block fading channel}
\acro{BG}[BG]{Bernoulli-Gaussian}
\acro{BGG}{Bernoulli-Generalized Gaussian}
\acro{BPAM}{binary pulse amplitude modulation}
\acro{BPDN}{Basis Pursuit Denoising}
\acro{BPPM}{binary pulse position modulation}
\acro{BPSK}{Binary Phase Shift Keying}
\acro{BPZF}{bandpass zonal filter}
\acro{BSC}{binary symmetric channels}              
\acro{BU}[BU]{Bernoulli-uniform}
\acro{BER}{bit error rate}
\acro{BS}{base station}
\acro{BW}{BandWidth}
\acro{BLLL}{ binary log-linear learning }

\acro{CP}{Cyclic Prefix}
\acrodef{cdf}[CDF]{cumulative distribution function}   
\acro{CDF}{Cumulative Distribution Function}
\acrodef{c.d.f.}[CDF]{cumulative distribution function}
\acro{CCDF}{complementary cumulative distribution function}
\acrodef{ccdf}[CCDF]{complementary CDF}               
\acrodef{c.c.d.f.}[CCDF]{complementary cumulative distribution function}
\acro{CD}{cooperative diversity}

\acro{CDMA}{Code Division Multiple Access}
\acro{ch.f.}{characteristic function}
\acro{CIR}{channel impulse response}
\acro{cosamp}[CoSaMP]{compressive sampling matching pursuit}
\acro{CR}{cognitive radio}
\acro{cs}[CS]{compressed sensing}                   
\acrodef{cscapital}[CS]{Compressed sensing} 
\acrodef{CS}[CS]{compressed sensing}
\acro{CSI}{channel state information}
\acro{CCSDS}{consultative committee for space data systems}
\acro{CC}{convolutional coding}
\acro{Covid19}[COVID-19]{Coronavirus disease}

\acro{DAA}{detect and avoid}
\acro{DAB}{digital audio broadcasting}
\acro{DCT}{discrete cosine transform}
\acro{dft}[DFT]{discrete Fourier transform}
\acro{DR}{distortion-rate}
\acro{DS}{direct sequence}
\acro{DS-SS}{direct-sequence spread-spectrum}
\acro{DTR}{differential transmitted-reference}
\acro{DVB-H}{digital video broadcasting\,--\,handheld}
\acro{DVB-T}{digital video broadcasting\,--\,terrestrial}
\acro{DL}{DownLink}
\acro{DSSS}{Direct Sequence Spread Spectrum}
\acro{DFT-s-OFDM}{Discrete Fourier Transform-spread-Orthogonal Frequency Division Multiplexing}
\acro{DAS}{Distributed Antenna System}
\acro{DNA}{DeoxyriboNucleic Acid}

\acro{EC}{European Commission}
\acro{EED}[EED]{exact eigenvalues distribution}
\acro{EIRP}{Equivalent Isotropically Radiated Power}
\acro{ELP}{equivalent low-pass}
\acro{eMBB}{Enhanced Mobile Broadband}
\acro{EMF}{ElectroMagnetic Field}
\acro{EU}{European union}
\acro{EI}{Exposure Index}
\acro{eICIC}{enhanced Inter-Cell Interference Coordination}

\acro{FC}[FC]{fusion center}
\acro{FCC}{Federal Communications Commission}
\acro{FEC}{forward error correction}
\acro{FFT}{fast Fourier transform}
\acro{FH}{frequency-hopping}
\acro{FH-SS}{frequency-hopping spread-spectrum}
\acrodef{FS}{Frame synchronization}
\acro{FSsmall}[FS]{frame synchronization}  
\acro{FDMA}{Frequency Division Multiple Access}

\acro{GA}{Gaussian approximation}
\acro{GF}{Galois field }
\acro{GG}{Generalized-Gaussian}
\acro{GIC}[GIC]{generalized information criterion}
\acro{GLRT}{generalized likelihood ratio test}
\acro{GPS}{Global Positioning System}
\acro{GMSK}{Gaussian Minimum Shift Keying}
\acro{GSMA}{Global System for Mobile communications Association}
\acro{GS}{ground station}
\acro{GMG}{ Grid-connected MicroGeneration}

\acro{HAP}{high altitude platform}
\acro{HetNet}{Heterogeneous network}

\acro{IDR}{information distortion-rate}
\acro{IFFT}{inverse fast Fourier transform}
\acro{iht}[IHT]{iterative hard thresholding}
\acro{i.i.d.}{independent, identically distributed}
\acro{IoT}{Internet of Things}                      
\acro{IR}{impulse radio}
\acro{lric}[LRIC]{lower restricted isometry constant}
\acro{lrict}[LRICt]{lower restricted isometry constant threshold}
\acro{ISI}{intersymbol interference}
\acro{ITU}{International Telecommunication Union}
\acro{ICNIRP}{International Commission on Non-Ionizing Radiation Protection}
\acro{IEEE}{Institute of Electrical and Electronics Engineers}
\acro{ICES}{IEEE international committee on electromagnetic safety}
\acro{IEC}{International Electrotechnical Commission}
\acro{IARC}{International Agency on Research on Cancer}
\acro{IS-95}{Interim Standard 95}

\acro{KPI}{Key Performance Indicator}

\acro{LEO}{low earth orbit}
\acro{LF}{likelihood function}
\acro{LLF}{log-likelihood function}
\acro{LLR}{log-likelihood ratio}
\acro{LLRT}{log-likelihood ratio test}
\acro{LoS}{Line-of-Sight}
\acro{LRT}{likelihood ratio test}
\acro{wlric}[LWRIC]{lower weak restricted isometry constant}
\acro{wlrict}[LWRICt]{LWRIC threshold}
\acro{LPWAN}{Low Power Wide Area Network}
\acro{LoRaWAN}{Low power long Range Wide Area Network}
\acro{NLoS}{Non-Line-of-Sight}
\acro{LiFi}[Li-Fi]{light-fidelity}
 \acro{LED}{light emitting diode}
 \acro{LABS}{LoS transmission with each ABS}
 \acro{NLABS}{NLoS transmission with each ABS}

\acro{MB}{multiband}
\acro{MC}{macro cell}
\acro{MDS}{mixed distributed source}
\acro{MF}{matched filter}
\acro{m.g.f.}{moment generating function}
\acro{MI}{mutual information}
\acro{MIMO}{Multiple-Input Multiple-Output}
\acro{MISO}{multiple-input single-output}
\acrodef{maxs}[MJSO]{maximum joint support cardinality}                       
\acro{ML}[ML]{maximum likelihood}
\acro{MMSE}{minimum mean-square error}
\acro{MMV}{multiple measurement vectors}
\acrodef{MOS}{model order selection}
\acro{M-PSK}[${M}$-PSK]{$M$-ary phase shift keying}                       
\acro{M-APSK}[${M}$-PSK]{$M$-ary asymmetric PSK} 
\acro{MP}{ multi-period}
\acro{MINLP}{mixed integer non-linear programming}

\acro{M-QAM}[$M$-QAM]{$M$-ary quadrature amplitude modulation}
\acro{MRC}{maximal ratio combiner}                  
\acro{maxs}[MSO]{maximum sparsity order}                                      
\acro{M2M}{Machine-to-Machine}                                                
\acro{MUI}{multi-user interference}
\acro{mMTC}{massive Machine Type Communications}      
\acro{mm-Wave}{millimeter-wave}
\acro{MP}{mobile phone}
\acro{MPE}{maximum permissible exposure}
\acro{MAC}{media access control}
\acro{NB}{narrowband}
\acro{NBI}{narrowband interference}
\acro{NLA}{nonlinear sparse approximation}
\acro{NLOS}{Non-Line of Sight}
\acro{NTIA}{National Telecommunications and Information Administration}
\acro{NTP}{National Toxicology Program}
\acro{NHS}{National Health Service}

\acro{LOS}{Line of Sight}

\acro{OC}{optimum combining}                             
\acro{OC}{optimum combining}
\acro{ODE}{operational distortion-energy}
\acro{ODR}{operational distortion-rate}
\acro{OFDM}{Orthogonal Frequency-Division Multiplexing}
\acro{omp}[OMP]{orthogonal matching pursuit}
\acro{OSMP}[OSMP]{orthogonal subspace matching pursuit}
\acro{OQAM}{offset quadrature amplitude modulation}
\acro{OQPSK}{offset QPSK}
\acro{OFDMA}{Orthogonal Frequency-division Multiple Access}
\acro{OPEX}{Operating Expenditures}
\acro{OQPSK/PM}{OQPSK with phase modulation}

\acro{PAM}{pulse amplitude modulation}
\acro{PAR}{peak-to-average ratio}
\acrodef{pdf}[PDF]{probability density function}                      
\acro{PDF}{probability density function}
\acrodef{p.d.f.}[PDF]{probability distribution function}
\acro{PDP}{power dispersion profile}
\acro{PMF}{probability mass function}                             
\acrodef{p.m.f.}[PMF]{probability mass function}
\acro{PN}{pseudo-noise}
\acro{PPM}{pulse position modulation}
\acro{PRake}{Partial Rake}
\acro{PSD}{power spectral density}
\acro{PSEP}{pairwise synchronization error probability}
\acro{PSK}{phase shift keying}
\acro{PD}{power density}
\acro{8-PSK}[$8$-PSK]{$8$-phase shift keying}
\acro{PPP}{Poisson point process}
\acro{PCP}{Poisson cluster process}
 
\acro{FSK}{Frequency Shift Keying}

\acro{QAM}{Quadrature Amplitude Modulation}
\acro{QPSK}{Quadrature Phase Shift Keying}
\acro{OQPSK/PM}{OQPSK with phase modulator }

\acro{RD}[RD]{raw data}
\acro{RDL}{"random data limit"}
\acro{ric}[RIC]{restricted isometry constant}
\acro{rict}[RICt]{restricted isometry constant threshold}
\acro{rip}[RIP]{restricted isometry property}
\acro{ROC}{receiver operating characteristic}
\acro{rq}[RQ]{Raleigh quotient}
\acro{RS}[RS]{Reed-Solomon}
\acro{RSC}[RSSC]{RS based source coding}
\acro{r.v.}{random variable}                               
\acro{R.V.}{random vector}
\acro{RMS}{root mean square}
\acro{RFR}{radiofrequency radiation}
\acro{RIS}{Reconfigurable Intelligent Surface}
\acro{RNA}{RiboNucleic Acid}
\acro{RRM}{Radio Resource Management}
\acro{RUE}{reference user equipments}
\acro{RAT}{radio access technology}
\acro{RB}{resource block}

\acro{SA}[SA-Music]{subspace-augmented MUSIC with OSMP}
\acro{SC}{small cell}
\acro{SCBSES}[SCBSES]{Source Compression Based Syndrome Encoding Scheme}
\acro{SCM}{sample covariance matrix}
\acro{SEP}{symbol error probability}
\acro{SG}[SG]{sparse-land Gaussian model}
\acro{SIMO}{single-input multiple-output}
\acro{SINR}{signal-to-interference plus noise ratio}
\acro{SIR}{signal-to-interference ratio}
\acro{SISO}{Single-Input Single-Output}
\acro{SMV}{single measurement vector}
\acro{SNR}[\textrm{SNR}]{signal-to-noise ratio} 
\acro{sp}[SP]{subspace pursuit}
\acro{SS}{spread spectrum}
\acro{SW}{sync word}
\acro{SAR}{specific absorption rate}
\acro{SSB}{synchronization signal block}
\acro{SR}{shrink and realign}

\acro{tUAV}{tethered Unmanned Aerial Vehicle}
\acro{TBS}{terrestrial base station}

\acro{uUAV}{untethered Unmanned Aerial Vehicle}
\acro{PDF}{probability density functions}

\acro{PL}{path-loss}

\acro{TH}{time-hopping}
\acro{ToA}{time-of-arrival}
\acro{TR}{transmitted-reference}
\acro{TW}{Tracy-Widom}
\acro{TWDT}{TW Distribution Tail}
\acro{TCM}{trellis coded modulation}
\acro{TDD}{Time-Division Duplexing}
\acro{TDMA}{Time Division Multiple Access}
\acro{Tx}{average transmit}

\acro{UAV}{Unmanned Aerial Vehicle}
\acro{uric}[URIC]{upper restricted isometry constant}
\acro{urict}[URICt]{upper restricted isometry constant threshold}
\acro{UWB}{ultrawide band}
\acro{UWBcap}[UWB]{Ultrawide band}   
\acro{URLLC}{Ultra Reliable Low Latency Communications}
         
\acro{wuric}[UWRIC]{upper weak restricted isometry constant}
\acro{wurict}[UWRICt]{UWRIC threshold}                
\acro{UE}{User Equipment}
\acro{UL}{UpLink}

\acro{WiM}[WiM]{weigh-in-motion}
\acro{WLAN}{wireless local area network}
\acro{wm}[WM]{Wishart matrix}                               
\acroplural{wm}[WM]{Wishart matrices}
\acro{WMAN}{wireless metropolitan area network}
\acro{WPAN}{wireless personal area network}
\acro{wric}[WRIC]{weak restricted isometry constant}
\acro{wrict}[WRICt]{weak restricted isometry constant thresholds}
\acro{wrip}[WRIP]{weak restricted isometry property}
\acro{WSN}{wireless sensor network}                        
\acro{WSS}{Wide-Sense Stationary}
\acro{WHO}{World Health Organization}
\acro{Wi-Fi}{Wireless Fidelity}

\acro{sss}[SpaSoSEnc]{sparse source syndrome encoding}

\acro{VLC}{Visible Light Communication}
\acro{VPN}{Virtual Private Network} 
\acro{RF}{Radio Frequency}
\acro{FSO}{Free Space Optics}
\acro{IoST}{Internet of Space Things}

\acro{GSM}{Global System for Mobile Communications}
\acro{2G}{Second-generation cellular network}
\acro{3G}{Third-generation cellular network}
\acro{4G}{Fourth-generation cellular network}
\acro{5G}{Fifth-generation cellular network}	
\acro{gNB}{next-generation Node-B Base Station}
\acro{NR}{New Radio}
\acro{UMTS}{Universal Mobile Telecommunications Service}
\acro{LTE}{Long Term Evolution}

\acro{QoS}{Quality of Service}
\end{acronym}
	
\newcommand{\SAR} {\mathrm{SAR}}
\newcommand{\WBSAR} {\mathrm{SAR}_{\mathsf{WB}}}
\newcommand{\gSAR} {\mathrm{SAR}_{10\si{\gram}}}
\newcommand{\Sab} {S_{\mathsf{ab}}}
\newcommand{\Eavg} {E_{\mathsf{avg}}}
\newcommand{\ft}{f_{\textsf{th}}}
\newcommand{\alphatf}{\alpha_{24}}

\title{
Evaluating the Accuracy of Stochastic Geometry Based Models for LEO Satellite Networks Analysis
}
\author{
Ruibo Wang, Mustafa A. Kishk, {\em Member, IEEE} and Mohamed-Slim Alouini, {\em Fellow, IEEE}
\thanks{Ruibo Wang and Mohamed-Slim Alouini are with King Abdullah University of Science and Technology (KAUST), CEMSE division, Thuwal 23955-6900, Saudi Arabia. Mustafa A. Kishk is with the Department of Electronic Engineering, National University of Ireland, Maynooth, W23 F2H6, Ireland. (e-mail: ruibo.wang@kaust.edu.sa; mustafa.kishk@mu.ie;
slim.alouini@kaust.edu.sa). 
}
\vspace{-8mm}
}
\maketitle

\begin{abstract}
This paper investigates the accuracy of recently proposed stochastic geometry-based modeling of low earth orbit (LEO) satellite networks. In particular, we use the Wasserstein Distance-inspired method to analyze the distances between different models, including Fibonacci lattice and orbit models. We propose an algorithm to calculate the distance between the generated point sets. Next, we test the algorithm's performance and analyze the distance between the stochastic geometry model and other more widely acceptable models using numerical results.
\end{abstract}


\section{Introduction}
With the recent advances in the low Earth orbit (LEO) satellites industry, a wide range of applications are anticipated to benefit from more than 4,700 LEO satellites launched in space \cite{yue2022security}. One main application is providing seamless global coverage and low-latency ultra-distance communication \cite{yaacoub2020key}. As a powerful mathematical tool, stochastic geometry is one of the few modeling methods that can provide analytical results for crucial performance metrics of satellite networks such as interference power, coverage probability, and latency \cite{haenggi2012stochastic,wang2022ultra}. Some literature has studied the coverage probability of satellite networks by modeling LEO satellite locations as Poisson point process (PPP) \cite{Al-1,Al-2}, binomial point process (BPP) \cite{talgat2020stochastic,ok-1,talgat2020nearest} and non-homogeneous Poisson point process (NPPP) \cite{ok-2}. Although stochastic geometry facilitate the analysis of the performance of satellite networks, the satellites rotate along fixed orbits in reality rather than follow the point process distribution on the entire spherical surface. Because of the difference between the stochastic geometry-based model and the actual orbit model, whether the stochastic geometry method is suitable for satellite network performance analysis has been the focus of discussion.
\par
Some literature tries to answer this question. The work in \cite{andrews2011tractable} indicates that the lower bound of coverage probability in the stochastic geometry-based model is as tight as the upper bound of the regular mesh model in a two-dimensional network. For a massive satellite constellation with a large number of orbits, \cite{ok-1} further proves that the coverage probability of the stochastic geometry-based model is consistent with that of the constellation with orbit model. However, the above literature focuses on a comparisons in terms of coverage probability, for which conclusions might not hold when other performance metrics are considered. In order to verify the accuracy of stochastic geometry based tools in the context of LEO satellite communications, a comparison with a general perspective is needed, but not the one focus on a specific performance metric. 
\par
The contributions of this letter can be summarized in the following three points. Firstly, we give expressions of the satellite orbit model distribution and spherical homogeneous BPP distribution, a common stochastic geometry model. To complete a comprehensive study, we considered geosynchronous devices, such as ground gateways and buoys on the sea. In many scenarios, they are suggested to be distributed uniformly to achieve better coverage or enhance availability when serving as relays. Therefore, the Fibonacci lattice model is introduced. The second contribution of this letter is that we establish the relationship between BPP and Fibonacci lattice, i.e., both of them provide solutions to the Tammas problem. Thirdly, this paper proposes a low-complexity algorithm to measure the accuracy of BPP when it is used to replace the other two models. What's more, the notations and corresponding descriptions involved in the following sections are shown in Table~\ref{table1}.

\section{Typical Models}
\label{section2}

\begin{table*}[]
\centering
\caption{Summary of Notations}\label{table1}
\begin{tabular}{M{3.5cm}|M{12.5cm}}
\hline
\textbf{Notation}                                            & \textbf{Description}        \\ \hline  \hline
$N_P$; $N_{\rm{orb}}$; $N_{\rm{NOI}}$ & Number of points in point process or point set; orbits; iterations.                               \\ \hline
$\theta_{\rm{BPP}}$, $\varphi_{\rm{BPP}}$; $\theta_{\rm{Fib}}$, $\varphi_{\rm{Fib}}$; $\theta_{\rm{orb}}$, $\varphi_{\rm{orb}}$          & Polar angle and azimuth angle of a point in Binomial point process; Fibonacci lattice-based point set; orbit model-based point process.   \\ \hline
$\theta_c$; $\theta_n$; $\gamma$ & Contact angle; nearest neighbor angle; inclination angle of orbit.      \\ \hline
$\Phi_o; \Phi_t$    & Positions of the original point process; the target point process or point set. \\ \hline
$v^I$; $v^C$, $v^{TC}$ &
Index vector of the original point   process; count vector, temporary count vector of the target point process. \\ \hline
$W_d$; $\Delta W_d$    & Distance between two point processes, distance increment.       \\ \hline \hline
\end{tabular} 
\end{table*}

\subsection{Homogeneous BPP}
BPP can be a better alternative to satellite networks because it is more suitable than PPP for modeling a finite number of points in a finite area. The following proposition gives a distribution of homogeneous BPP on a sphere.

\begin{proposition} \label{proposition1}
For a point in homogeneous BPP, the azimuth angle is uniformly distributed between 0 and $2\pi$, i.e $\varphi_{\rm{BPP}} \sim \mathcal{U}[0,2\pi]$, and the \ac{CDF} of each point's polar angle (of the spherical coordinate) $\theta_{\rm{BPP}}$ follows
\begin{equation}
    F_{\theta_{\rm{BPP}}}\left( \theta \right) = \frac{1-\cos\theta}{2}, \ 0 \leq \theta_{\rm{BPP}} \leq \pi,
\end{equation}
and $\theta_{\rm{BPP}}$ can be generated by,
\begin{equation}\label{generate_BPP}
    \theta_{\rm{BPP}} = \arccos\left(1-2\, \mathcal{U}[0,1]\right), \ 0 \leq \theta_{\rm{BPP}} \leq \pi.
\end{equation}
\begin{proof}
See Appendix~\ref{app:proposition1}.
\end{proof}
\end{proposition}
Note that the BPP given in subsequent parts of this letter means homogeneous BPP unless otherwise stated.

\begin{figure}[h]
	\centering
	\includegraphics[width=0.98\linewidth]{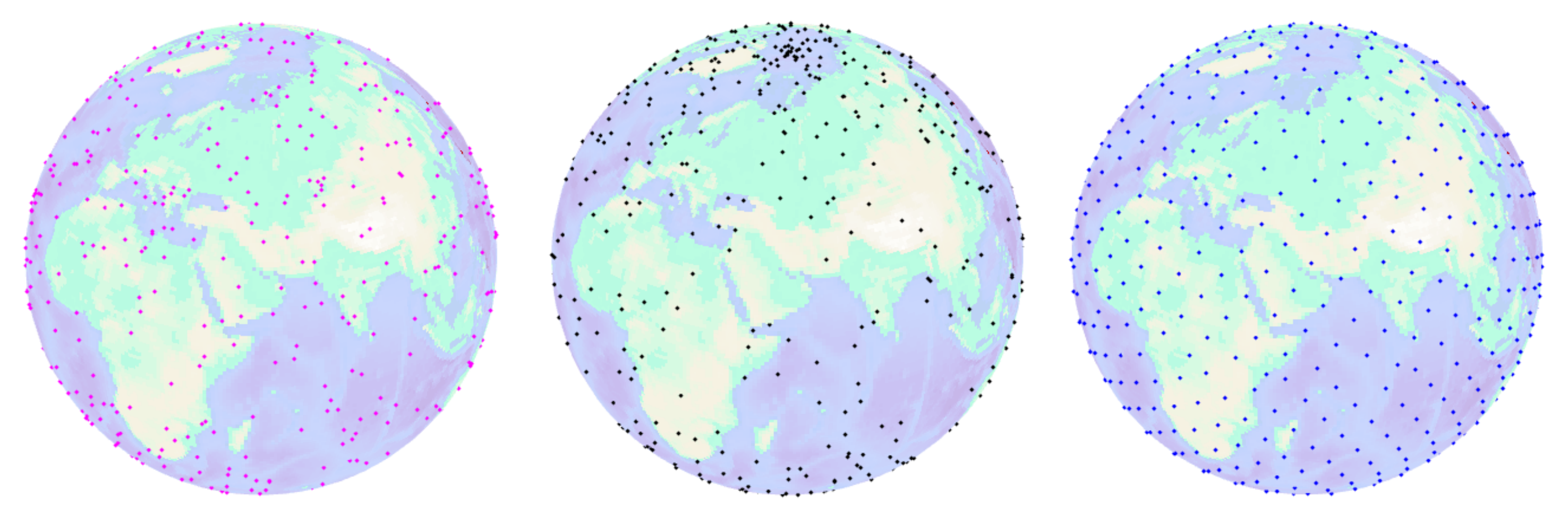}
    	\caption{Comparison of homogeneous BPP, non-homogeneous BPP and Fibonacci lattice model-based point set.}
	\label{fig:Figure1}
\end{figure}

\subsection{Fibonacci Lattice-Based Point Set}
As shown in the right side of Fig.~\ref{fig:Figure1}, the point set based on the Fibonacci lattice is stationary to the Earth and highly uniform. The above distribution can be obtained by solving Tammes problem \cite{erber1991equilibrium}, which is a problem in packing a given number of points on a sphere such that the minimum distance between points is maximized,
\begin{equation}\label{P_1}
    \mathscr{P}_1:\quad d_{\rm{opt}} = \underset{\mathrm{All \, of \, the} \, x_i \in \Phi_t} {\rm{maximize}} \, \min_{i\neq j}\{d_P(x_i,x_j)\},
\end{equation}
where $x_i\left(R, \theta_i, \varphi_i \right)$ is the position of point $i$, for gateways and buoys, $R=6371$km is the radius of the Earth, $\theta_i$ and $\phi_i$ are polar angle and azimuth angle of point $i$. $\Phi_t = \{ x_1, x_2, \dots ,x_{N_P} \}$ is the point set and  $d_P(x_i,x_j)$ in (\ref{P_1}) is the distance between $x_i$ and $x_j$, which can be expressed as, 
\begin{sequation}
\begin{split}
\label{d_P}
    & d_P(x_i,x_j)  \\
    & = R \sqrt{2\Big(1 - \cos{\theta_i}\cos{\theta_j}-\sin{\theta_i}\sin{\theta_j}\cos\left(\varphi_i-\varphi_j\right)\Big)}.
\end{split}
\end{sequation}
To obtain $\Phi_t$ from the above optimization problem, a well-studied point set called the Fibonacci lattice is given in the following proposition \cite{gonzalez2010measurement}.

\begin{proposition}
For the $i^{th}$ point in Fibonacci lattice-based point set, the polar angle $\theta_{\rm{Fib},i}$ and azimuth angle $\varphi_{\rm{Fib},i}$ are calculated by \cite{gonzalez2010measurement},
\begin{sequation}
\begin{split}
\left\{\begin{matrix}
\theta_{\rm{Fib},i}=\arccos \left ( {\frac{2i-1}{N_P-1}} \right ),\\ 
\varphi_{\rm{Fib},i}=\left ( \sqrt{5}-1 \right ) \pi i,
\end{matrix}\right. \, \mathrm{when} \ & i \leq  \left \lceil \frac{N_P}{2} \right \rceil,
\\
\left\{\begin{matrix}
\theta_{\rm{Fib},i}=\pi - \arccos \left ( {\frac{2\left( i- \left \lceil \frac{N_P}{2} \right \rceil\right )-1}{N_P-1}} \right ),\\
\varphi_{\rm{Fib},i}=\left ( \sqrt{5}-1 \right ) \pi \cdot \left( i- \left \lceil \frac{N_P}{2} \right \rceil\right ),
\end{matrix}\right. & \, \mathrm{when} \ i > \left \lceil \frac{N_P}{2} \right \rceil.
\end{split}
\end{sequation}
where $N_P$ is the number of points in point set $\Phi_t$, and $\left \lceil \cdot \right \rceil$ represents round up to an integer.
\end{proposition}

Based on the nearest neighbor angle of BPP, we give an approximate solution to Tammes problem.

\begin{corollary}\label{corollary2}
When packing $N_P$ points on a sphere with radius $R$, the maximum value $d_{\rm{opt}}$ of the minimum distance between points can be approximate as,
\begin{equation}
    d_{\rm{opt}} \approx 2R \sin\left( \pi \prod \limits_{i=1}^{N_P-1} \frac{2i-1}{2i} \right).
\end{equation}
\begin{proof}
See Appendix~\ref{app:corollary2}.
\end{proof}
\end{corollary}

\subsection{Orbit Model-Based Point Process}
\begin{figure}[h]
	\centering
	\includegraphics[width=0.85\linewidth]{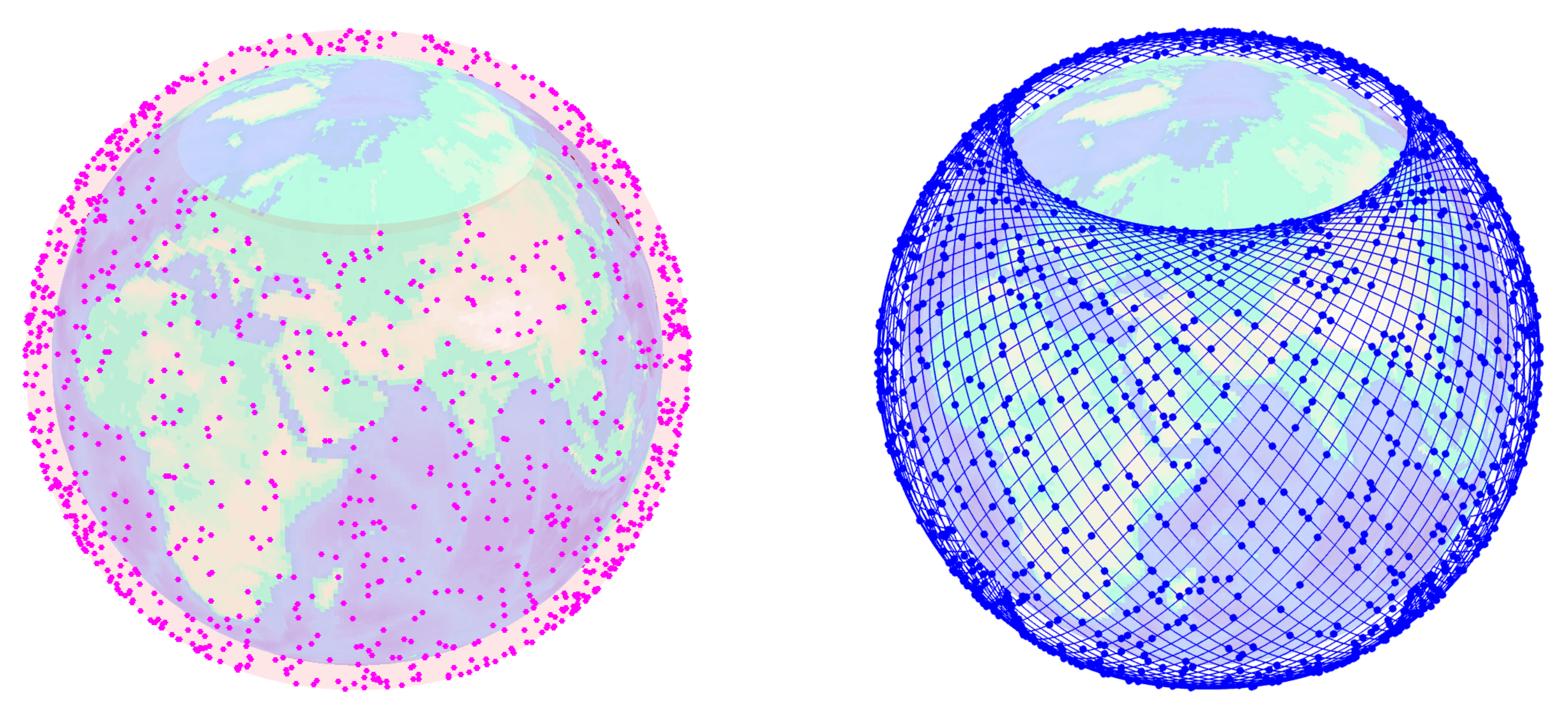}
	\caption{Comparison of homogeneous BPP and orbit model-based point process.}
	\label{fig:Figure2}
\end{figure}

The orbit model of part of the Starlink constellation at 550km is shown on the right side of Fig.~\ref{fig:Figure2}. It consists of 72 isometric orbits, each uniformly distributed with 22 satellites, with an orbital inclination of 53 degrees \cite{hu2020directed}. For increasing coverage and avoiding collisions, the satellites in each orbit are also assumed to be uniformly distributed. Since LEO satellites are not geostationary, they are modeled as a homogeneous stochastic point process in fixed orbits. The following proposition gives the realization of the above point process.

\begin{proposition}\label{proposition2}
For a point in orbit model-based point process, the CDF of its polar angle $F_{\theta_{\rm{orb}}}\left(\theta\right)$ is given by, 
\begin{equation}
    F_{\theta_{\rm{orb}}}\left(\theta\right) = \frac{\cos\gamma - \cos\theta}{2\cos\gamma}, \ \gamma \leq \theta \leq \pi-\gamma,
\end{equation}
where $\gamma$ is the orbit inclination. For a given $\theta_{\rm{orb}}$, its azimuth angle $\varphi_{\rm{orb}}$ follows a discrete uniform distribution,
\begin{equation}
\begin{split}
    \mathbbm{P}\left[\varphi_{\rm{orb}}=\frac{2k\pi}{N_{\rm{orb}}}\pm \arcsin\left(\frac{\tan\theta_{\rm{orb}}}{\tan\gamma}\right)\right] = \frac{1}{2N_{\rm{orb}}}&,\\ 
    k=1,2,...,N_{\rm{orb}}&,
\end{split}
\end{equation}
where $N_{\rm{orb}}$ is the number of orbits. 
\begin{proof}
See Appendix~\ref{app:proposition2}.
\end{proof}
\end{proposition}

\section{Wasserstein Distance-Inspired Difference Measurement} \label{section3}

Based on the concept of Wasserstein distance, we calculate the distance between two point processes or between a point process and a set of points. For discrete distributions, the squared of Wasserstein distance is the minimum energy required to move one distribution to another. Similarly, we sample the original process $\Phi_o$ (for BPP) and the target point process/set $\Phi_t$ (for Fibonacci lattice-based point set or orbit model-based point process) to obtain the coordinates $\Phi_o=\{x_1,x_2,\dots,x_{N_p}\}$ and $\Phi_t=\{\widetilde{x}_1,\widetilde{x}_2,\dots,\widetilde{x}_{N_p}\}$ respectively. Note that the number of points in $\Phi_o$ and $\Phi_t$ must be equal. To measure the distance between $\Phi_o$ and $\Phi_t$, we calculate the minimum power required to move from $\Phi_o$ to $\Phi_t$. The corresponding optimization problem is expressed as follows,

\begin{subequations} 
	\begin{alignat}{2}
		\mathscr{P}_2:\quad W_d = \ &\underset{F}{\min}  & \ &  \sqrt{\sum_{i=1}^{N_P}\sum_{j=1}^{N_P} F_{ij} \cdot d_P^2\left( x_i, \widetilde{x}_j \right)},\\
		&\textrm{s.t.}    &      & \sum_{j=1}^{N_P} F_{ij} = 1,\,\forall i,\\
		&                  &      &   \sum_{i=1}^{N_P} F_{ij} = 1,\,\forall j,\\
		&                  &      &   F_{ij} = 0 \ \mathrm{or} \ F_{ij} = 1, \ \forall i,j.
	\end{alignat}
\end{subequations}
where $d_P\left( x_i, \widetilde{x}_j \right)$ is defined in (\ref{d_P}), $F$ is a permutation matrix which represents a moving scheme. According to the whole permutation formula, there are $N_P!$ moving schemes, which brings in a huge computing burden. 
\par
Therefore, we propose the sampling algorithm for the distance with the following steps: ({\romannumeral1}) Let the points in $\Phi_o$ associate to their nearest points in $\Phi_t$; ({\romannumeral2}) Traverse points in $\Phi_t$ which have been associated multiple times, left the nearest point in $\Phi_o$ and disassociated the others; ({\romannumeral3}) Repeat the above process for the disassociated points in $\Phi_o$ and the unassociated points in $\Phi_t$, until all the one-to-one association is achieved; ({\romannumeral4}) Move $\Phi_o$ to $\Phi_t$ according to the association and calculate $W_d$. According to the above analysis, each round's computation amount of step ({\romannumeral1}) and ({\romannumeral2}) is no more than that of a moving scheme in $\mathscr{P}_2$. Since at least one point in $\Phi_t$ is associated in each round, the number of rounds will not exceed $N_P$. Compared with $N_P!$ moving scheme in $\mathscr{P}_2$, the algorithm~\ref{alg.Algorithm} has low computational complexity.

 \begin{algorithm}[!ht] 
	\caption{Sampling Algorithm for the Distance}
	\label{alg.Algorithm} 
	\begin{algorithmic} [1]
		
		\STATE \textbf{Input}: Positions of original point processes $\Phi_o$, positions of target point set/process $\Phi_t$, and the number of iterations $N_{\rm{NOI}}$.
		
		\STATE \textbf{Initialize}: Initialize index vector $v^I \leftarrow \mathbf{0}_{N_P \times 1}$, count vector $v^C \leftarrow \mathbf{0}_{N_P \times 1}$, and distance matrix $D_{ij} \leftarrow d_P\left(x_i,\widetilde{x}_j\right)$ for all $i, j \leq N_P$.
		
		\FOR{$n = 1 : N_{\rm{NOI}}$}
		
		\WHILE{$\exists \, v_j^C \neq 1, j = 1,2,\cdots,N_P$}
		
		\STATE Perform a round of association according to algorithm~\ref{alg.Iteration}, renew $v^I,v^C$, and $W_d \leftarrow \sqrt{W_d^2 + \Delta W_d^2}$.

		\ENDWHILE
		
		\ENDFOR
		\STATE $W_d \leftarrow \frac{W_d}{N_{\rm{NOI}}}$.
		\STATE \textbf{Output}: Distance between point set/processes $W_d$.
	\end{algorithmic}
\end{algorithm}	
In the above algorithm, $v_i^I$ is used to record the index of the point in $\Phi_t$ which is associated to point $x_i$ in $\Phi_o$, and $v_j^C$ is used to record how many points are associated with point $\widetilde{x}_j$ in $\Phi_t$. Finally, a round of association for the sampling algorithm is given in algorithm \ref{alg.Iteration}.

 \begin{algorithm}[!ht] 
	\caption{A Round of Association for Sampling Algorithm}
	\label{alg.Iteration} 
	\begin{algorithmic} [1]
		
		\STATE \textbf{Input}: Distance matrix $D$, index vector $v^I$, and count vector $v^C$.
		
		\STATE \textbf{Initialize}: Distance increment $\Delta W_d \leftarrow 0 $, temporary count vector $v^{TC} \leftarrow \mathbf{0}_{N_P \times 1}$.
		
        \FOR{$i = 1 : N_P$}
		\IF{$v_i^I = 0$}
		\STATE  $j_{\rm{opt}} \leftarrow \underset{j, \, \rm{for \, all} \, v_j^C=0}{\arg\min} D_{ij}$.
		\STATE $v_i^I \leftarrow j_{\rm{opt}}$.
		\STATE $v_{j_{\rm{opt}}}^{TC} \leftarrow v_{j_{\rm{opt}}}^{TC} + 1 $.
		\ENDIF
		\ENDFOR
		
		\FOR{$j = 1 : N_P$}
		\IF{$v_j^{TC} = 1$}
		\STATE $v_j^{T} \leftarrow 1$.
		\STATE $\Delta W_d \leftarrow \sqrt{\Delta W_d^2 + D_{i_{\rm{opt}}j}^2}$, where $i_{\rm{opt}}$ is derived by $v_i^I = j$.
		
		\ELSIF {$v_j^{TC} > 1$}
		\STATE $v_j^{T} \leftarrow 1$.
		\STATE $i_{\rm{opt}} \leftarrow \underset{i, \, \rm{for \, all} \, v_i^I=j}{\arg\min} D_{ij}$.
		\STATE $\Delta W_d \leftarrow \sqrt{\Delta W_d^2 + D_{i_{\rm{opt}}j}^2}$.
		\STATE $v_i^I \leftarrow 0$, for all $v_i^I=j$ and $i\neq i_{\rm{opt}}$.
		\ENDIF
		\ENDFOR
		
		\STATE \textbf{Output}: Distance increment $\Delta W_d$, index vector $v^I$, and count vector $v^C$.
	\end{algorithmic}
\end{algorithm}

\section{Numerical Results} \label{section4}

\begin{figure}[h]
	\centering
	\includegraphics[width=0.7\linewidth]{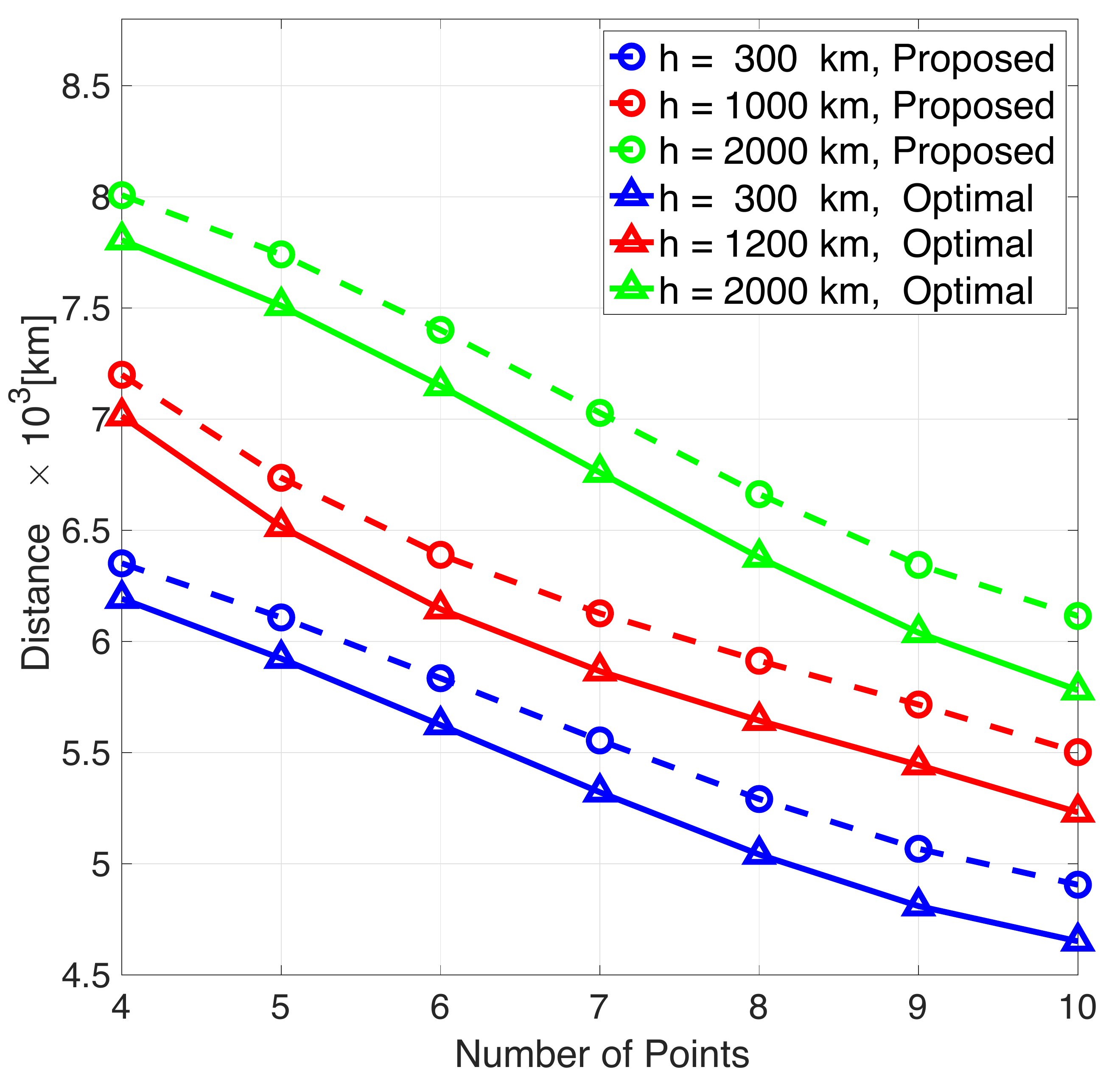}
	\caption{Optimality analysis of the proposed algorithm.}
    \label{fig:Figure3}
\end{figure}
\begin{figure}[h]
	\centering
	\includegraphics[width=0.7\linewidth]{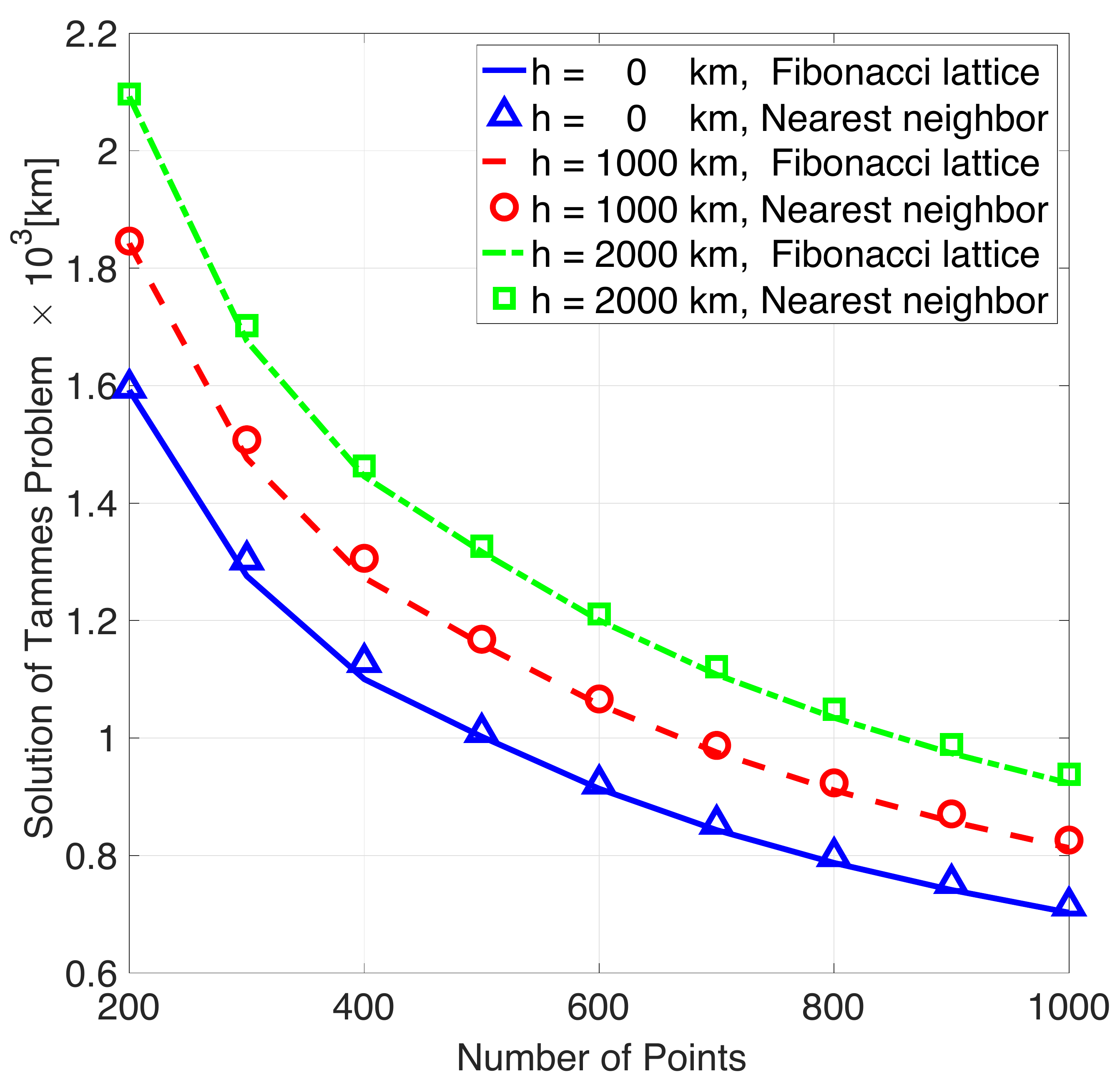}
	\caption{Verification of the accuracy in Corollary~\ref{corollary2}.}
    \label{fig:Figure4}
\end{figure}
\begin{figure}[h]
	\centering
	\includegraphics[width=0.7\linewidth]{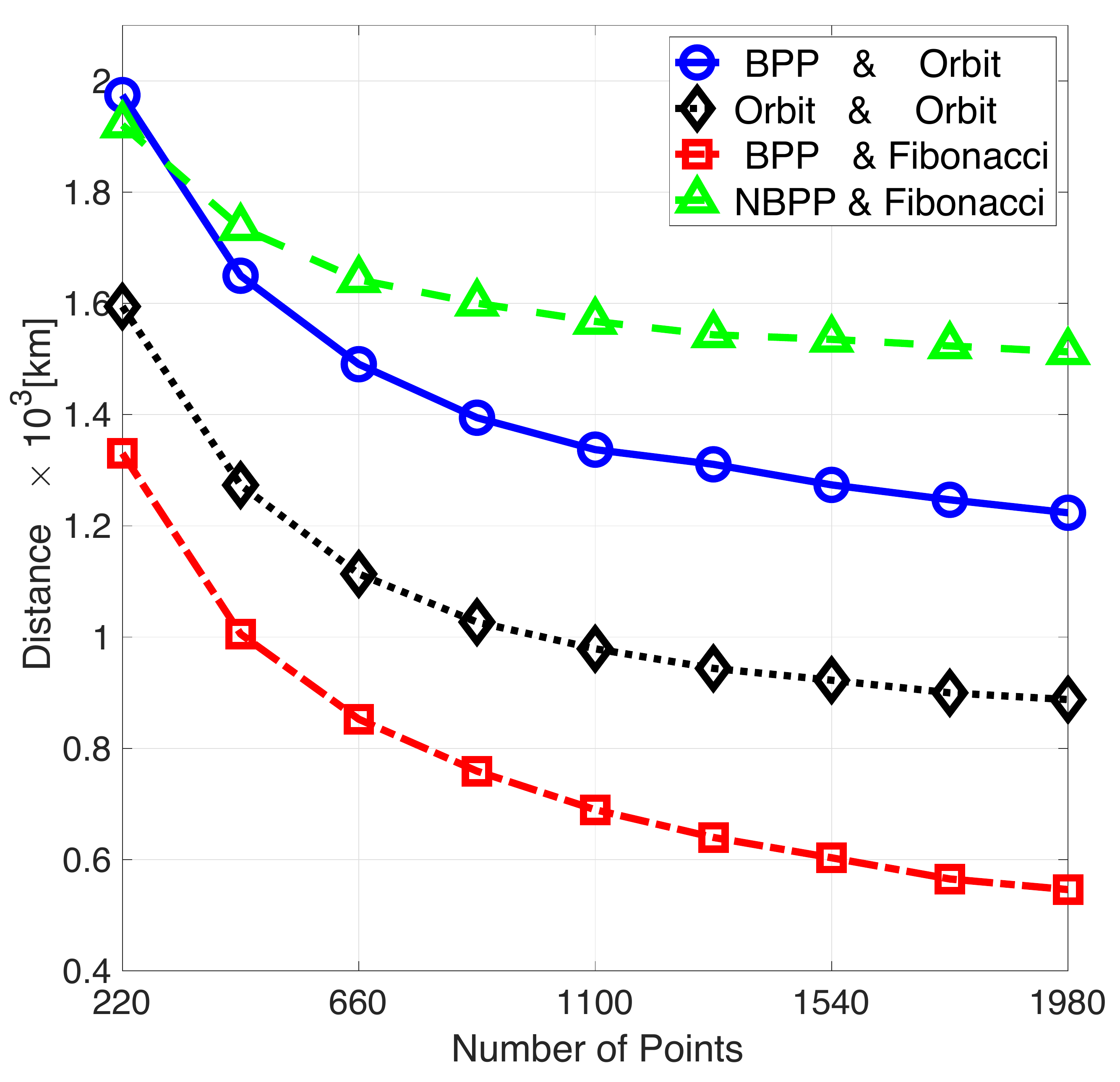}
	\caption{Figure of the distances between point processes/sets.}
    \label{fig:Figure5}
\end{figure}
\begin{figure}[h]
	\centering
	\includegraphics[width=0.7\linewidth]{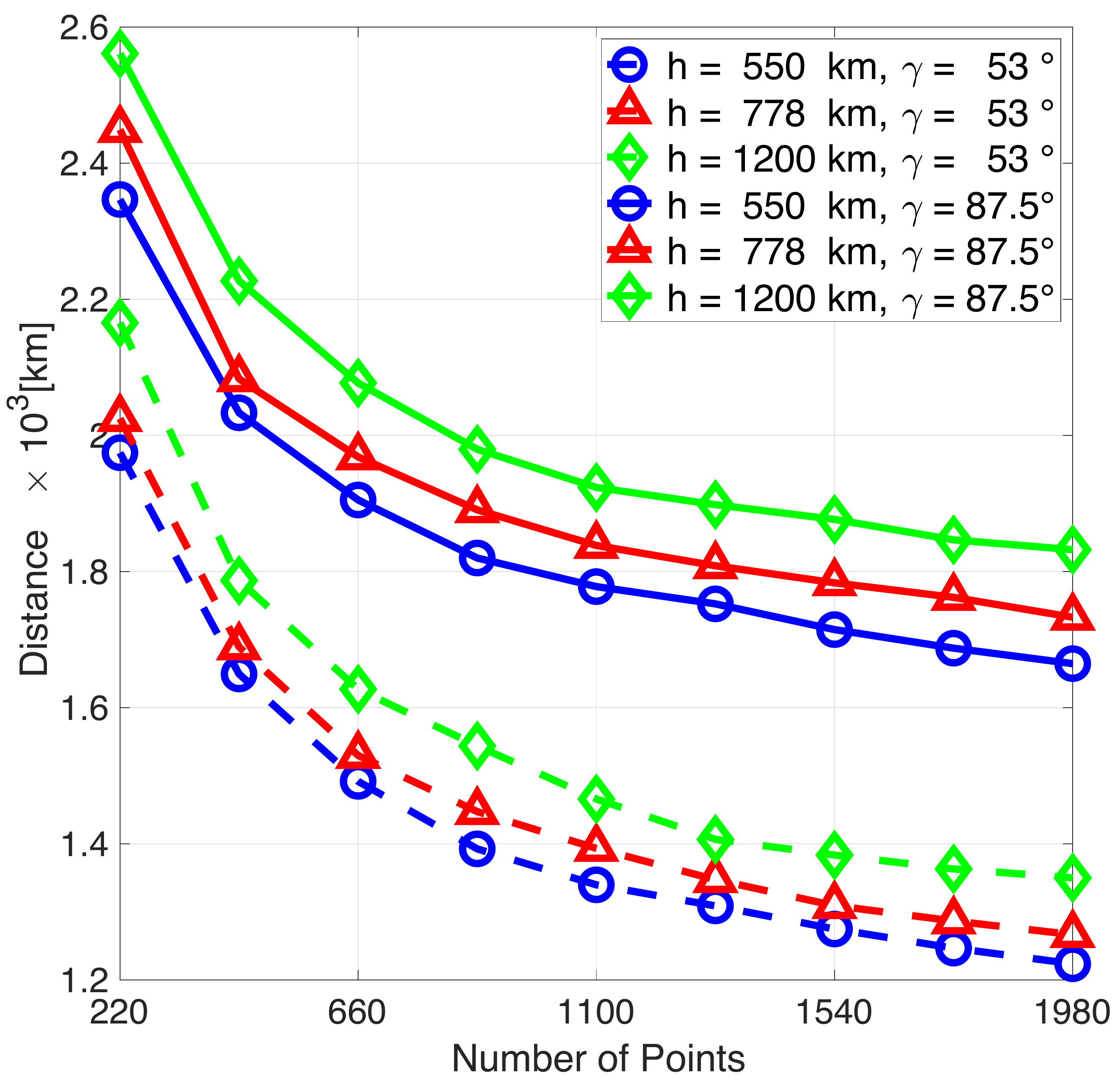}
	\caption{The analysis of the distance between BPP and orbit model-based point processes.}
	\label{fig:Figure6}
\end{figure}

In this section, we begin with the verification process in Fig.~\ref{fig:Figure3} and \ref{fig:Figure4}, then the numerical results obtained by the algorithm~\ref{alg.Algorithm} are analyzed in Fig.~\ref{fig:Figure5} and \ref{fig:Figure6}. We run $N_{\rm{NOI}}=10^6$ iterations for each point in the figures and take the mean of the obtained $10^6$ distances as the final result. In each iteration, the coordinates of points will be reconstructed for BPP or orbit model-based point process. 

In Fig.~\ref{fig:Figure3}, the optimality of algorithm~\ref{alg.Algorithm} is verified by comparing the distance between BPP and orbit model-based point process. $h=R-6371$km is the height of the satellite constellation, where $ R $ is the radius of the sphere over which the satellites are located. The solid line is obtained by traversing all schemes $F$ in $\mathscr{P}_2$, which is optimal. From Fig.~\ref{fig:Figure3}, we notice that the distances obtained by the algorithm (the dashed lines) is not far from the optimal distance, even when there are less than 10 points. In Fig.~\ref{fig:Figure4}, we verify accuracy of  corollary~\ref{corollary2}. Since the points and lines fit well, we prove that the results in Corollary~\ref{corollary2} based on the nearest neighbor angle can provide an excellent approximation to the Tammes problem at any height. 

Fig.~\ref{fig:Figure5} and Fig.~\ref{fig:Figure6} analyze the distances between BPP and other two models. To show the uniformity of BPP obtained in proposition~\ref{proposition1}, we design a non-homogeneous binomial point process (NBPP) shown in the middle sub-graph of Fig.~\ref{fig:Figure1} as a comparison. Both azimuth and polar angles of NBPP follow uniform distributions, i.e., $\varphi_{\rm{NBPP}} \sim \mathcal{U}[0,2\pi]$ and $\theta_{\rm{NBPP}} \sim \mathcal{U}[0,\pi]$. For the red and green lines of Fig.~\ref{fig:Figure5} and subfigures in Fig.~\ref{fig:Figure1}, Fibonacci lattice-based point set, BPP, and NBPP are located on the ground ($h=0$). The results of these two lines show that BPP is more suitable for substituting the Fibonacci lattice-based point set than NBPP. When the number of points exceeds 1000, the distance between BPP and Fibonacci lattice-based point set is small enough. 

\par
In the black and blue lines of Fig.~\ref{fig:Figure5}, $h=550$km and $\gamma=53^{\circ}$. For the orbit model-based point process in Fig.~\ref{fig:Figure5} and Fig.~\ref{fig:Figure6}, the number of satellites in each orbit is fixed at 22. As shown in Fig.~\ref{fig:Figure5}, since the orbit model generates a stochastic point process, there are also distances between samples at different times. However, the distance between BPP and orbit model-based point process is not much greater than the distance of this process itself. Therefore, the distance caused by substituting the orbit model with a stochastic geometry model is acceptable. Fig.~\ref{fig:Figure6} compares the distance between BPP and orbit model-based point processes under different parameters. Smaller orbit inclinations, lower altitudes and more satellites lead to smaller distances. Stochastic geometry model is more suitable in analyzing Starlink mega-constellation ($\gamma \approx 53^{\circ}, N_P=41927, h=550\rm{km}$), than Iridium ($\gamma \approx 87.5^{\circ}, N_P=81, h=778\rm{km}$) and OneWeb ($\gamma \approx 87.5^{\circ}, N_P=720, h=1200\rm{km}$) constellations \cite{yue2022security}.

\section{Conclusion}
As a point process suitable for theoretical analysis, homogeneous BPP is an excellent substitute for both Fibonacci lattice-based point set and orbit model-based point process. According to the results obtained by the proposed algorithm which is used to measure the distance between point processes, BPP has a better substitution effect for a point process/set with a lower altitude and larger number of points, such as a point process based on a massive LEO satellite constellation.
\par

\appendices
\section{Proof of Proposition~\ref{proposition1}}\label{app:proposition1}

First, we introduce a well-known distribution in stochastic geometry called the contact angle distribution. The contact angle of a reference point is the dome angle between the point and its nearest satellite, and the dome angle is the connection between the reference point and the center of the earth and the connection between the satellite to the center of the Earth \cite{wang2022stochastic}. 
The \ac{CDF} of the contact angle $F_{\theta_c}\left(\theta\right)$ is given by,
\begin{equation}\label{CDF of contact}
    F_{\theta_c}\left(\theta\right) = 1 - \left( \frac{1+\cos \theta}{2} \right) ^ {N_P}, 0 \leq \theta_c \leq \pi,
\end{equation}
where $N_P$ is the number of satellites. The CDF of each point's polar angle is  $F_{\theta_c}\left(\theta\right)$ at $N_P=1$. Let the CDF of the polar angle be equal to a uniformly distributed random variable on $[0,1]$, i.e. $F_{\theta_{\rm{BPP}}}\left( \theta \right) = \frac{1-\cos\theta}{2} = \mathcal{U}[0,1]$, the result in (\ref{generate_BPP}) can be derived.

\section{Proof of Corollary~\ref{corollary2}}\label{app:corollary2}
We start the proof from the concept of the nearest neighbor angle. The nearest neighbor angle is the dome angle between a reference point from the point process and its nearest neighbor (a gateway or buoy). By definition, the CDF of $F_{\theta_n}\left(\theta\right)$ nearest neighbor angle can be derived directly from (\ref{CDF of contact}),
\begin{equation}
    F_{\theta_n}\left(\theta\right) = 1 - \left( \frac{1+\cos \theta}{2} \right) ^ {N_P-1}.
\end{equation}
Take the expectation of $\theta_n$, we have
\begin{sequation}\label{appen1}
\begin{split}
\mathbb{E}\left[\theta_n\right] &= \int_0^\pi 1 - F_{\theta_n}\left(\theta\right)\mathrm{d}\theta = \int_0^\pi \left(\frac{1+\cos\theta}{2}\right)^{N_P-1}
\mathrm{d}\theta\\
& = 2\int_0^{\frac\pi{2}}\left(\cos\widetilde{\theta} \right)^{2N_P-2} \mathrm{d}\widetilde{\theta} \overset{(a)}{\approx} \pi \prod \limits_{i=1}^{N_P-1} \frac{2i-1}{2i},
\end{split}
\end{sequation}
where (a) follows Wallis' integrals. In the isosceles triangle formed by the reference point, its nearest neighbor, and the center of the earth, the simple geometric relationship between the average dome angle $\mathbb{E}\left[\theta_n\right]$ and the average distance $\overline{d}_n$ between the reference point and its nearest neighbor is obtained,
\begin{equation}\label{appen2}
\overline{d}_n = 2R \sin\left( \mathbb{E}\left[\theta_n\right] \right).
\end{equation}
$\overline{d}_n$ can be regarded as an approximation of $d_{\rm{opt}}$. Substituted (\ref{appen1}) into (\ref{appen2}),  corollary~\ref{corollary2} is proved.

\section{Proof of Proposition~\ref{proposition2}}\label{app:proposition2}
Since the trajectory of the satellite is continuous, so is the CDF of the polar angle $F_{\theta_{\rm{orb}}}\left(\theta\right)$. $F_{\theta_{\rm{orb}}}\left(\theta\right)$ can be obtained by the normalization of the interval $\gamma \leq \theta_{\rm{BPP}} \leq \pi - \gamma$ in $F_{\theta_{\rm{BPP}}}\left(\theta\right)$,

\begin{equation}
\begin{split}
    F_{\theta_{\rm{orb}}}\left(\theta\right) & = \frac{F_{\theta_{\rm{BPP}}}\left(\theta\right)-F_{\theta_{\rm{BPP}}}\left(\gamma\right)}{F_{\theta_{\rm{BPP}}}\left(\pi-\gamma\right)-F_{\theta_{\rm{BPP}}}\left(\gamma\right)} \\
    & = \frac{\cos\gamma - \cos\theta}{\cos\gamma - \cos\left(\pi-\gamma\right)} = \frac{\cos\gamma - \cos\theta}{2\cos\gamma}.
\end{split}
\end{equation}

Since the orbit is discretized, the azimuth angle is given by a discrete uniform distribution. First, we derive a particular orbital equation and establish the connection between $\theta_{\rm{orb}}$ and $\varphi_{\rm{orb}}$. For a circle passing point $(R,\frac{\pi}{2},\gamma)$ with its center at the center of the earth, the following equation is satisfied,
\begin{equation}
    \tan\gamma\cos\theta_{\rm{orb}}\sin\varphi_{\rm{orb}}-\sin\theta_{\rm{orb}}=0,
\end{equation}
and there are two solutions for azimuth angles,
\begin{equation}
\begin{split}
    \varphi_{\rm{orb}} &=  \pm \arcsin\left(\frac{\tan\theta_{\rm{orb}}}{\tan\gamma}\right).
\end{split}
\end{equation}
By changing the coordinates of the points which the circle passes, the relationship between $\theta_{\rm{orb}}$ and $\varphi_{\rm{orb}}$ in the rest of the orbits can be represented. Finally, since the probability of a satellite appearing in each orbit is equal, the proposition is proved.



\bibliographystyle{IEEEtran}
\bibliography{references}

\end{document}